\documentclass[prb,10pt,twocolumn,superscriptaddress]{revtex4-2}

\usepackage{graphicx}

\usepackage{slashed}

\begin{document}

\title{Many-body theory of radiative lifetimes of exciton-trion superposition states in doped two-dimensional materials}

\author{Farhan Rana}
\address{School of Electrical and Computer Engineering, Cornell University, Ithaca, NY 14853}
\author{Okan Koksal}
\address{School of Electrical and Computer Engineering, Cornell University, Ithaca, NY 14853}
\author{Minwoo Jung}
\address{School of Applied and Engineering Physics, Cornell University, Ithaca, NY 14853}
\author{Gennady Shvets}
\address{School of Applied and Engineering Physics, Cornell University, Ithaca, NY 14853}
\author{Christina Manolatou}
\address{School of Electrical and Computer Engineering, Cornell University, Ithaca, NY 14853}
\email{fr37@cornell.edu}

\begin{abstract}
  Optical absorption and emission spectra of doped two-dimensional (2D) materials exhibit sharp peaks that are often mistakenly identified with pure excitons and pure trions (or charged excitons), but both peaks have been recently attributed to superpositions of 2-body exciton and 4-body trion states and correspond to the approximate energy eigenstates in doped 2D materials. In this paper, we present the radiative lifetimes of these exciton-trion superposition energy eigenstates using a many-body formalism that is appropriate given the many-body nature of the strongly coupled exciton and trion states in doped 2D materials. Whereas the exciton component of these superposition eigenstates are optically coupled to the material ground state, and can emit a photon and decay into the material ground state provided the momentum of the eigenstate is within the light cone, the trion component is optically coupled only to the excited states of the material and can emit a photon even when the momentum of the eigenstate is outside the light cone. In an electron-doped 2D material, when a 4-body trion state with momentum outside the light cone recombines radiatively, and a photon is emitted with a momentum inside the light cone, the excess momentum is taken by an electron-hole pair left behind in the conduction band. The radiative lifetimes of the exciton-trion superposition states, with momenta inside the light cone, are found to be in the few hundred femtoseconds to a few picoseconds range and are strong functions of the doping density. The radiative lifetimes of exciton-trion superposition states, with momenta outside the light cone, are in the few hundred picoseconds to a few nanoseconds range and are again strongly dependent on the doping density. The doping density dependence of the radiative lifetimes of the two peaks in the optical emission spectra follows the doping density dependence of the spectral weights of the same two peaks observed in the optical absorption spectra as both have their origins in the Coulomb coupling between the excitons and trions in doped 2D materials.
\end{abstract}  

\maketitle

\section{Introduction}

Optical absorption and emission spectra of doped two-dimensional (2D) materials in general, and of transition metal dichalcogenides (TMDs) in particular, exhibit sharp and distinct peaks that are often attributed to neutral and charged excitons (or trions)~\cite{Fai13, Changjian14, Berk13, Chernikov14, Chernikov15,Combes03,Combes12,Suris01,Suris01b,Urba17,Kheng93}. Although optical signatures of excitons and trions in doped semiconductors have been observed for a long time~\cite{Kheng93}, their nature, especially of trions, in doped materials had remained somewhat of a mystery. For one, it was difficult to understand how a photon, being a boson, could get absorbed and create a trion, if a trion is taken to be fermionic bound state of three particles. Second, it was not clear what happened to one of the charged particles left behind when a trion emitted a photon. Pauli's exclusion required the left behind charged particle to be deposited outside the Fermi sea, but the energy and momentum conservation requirements following from Pauli's exclusion were never observed in the measured photoluminescence spectra. Third, the variation of the energy separation of the two peaks observed in the optical absorption spectra, as well as the spectral weight transfer between these two peaks with doping, did not seem to follow from the assumption of excitons and trions being independent excitations. 

\begin{figure}
  \begin{center}
   \includegraphics[width=1.0\columnwidth]{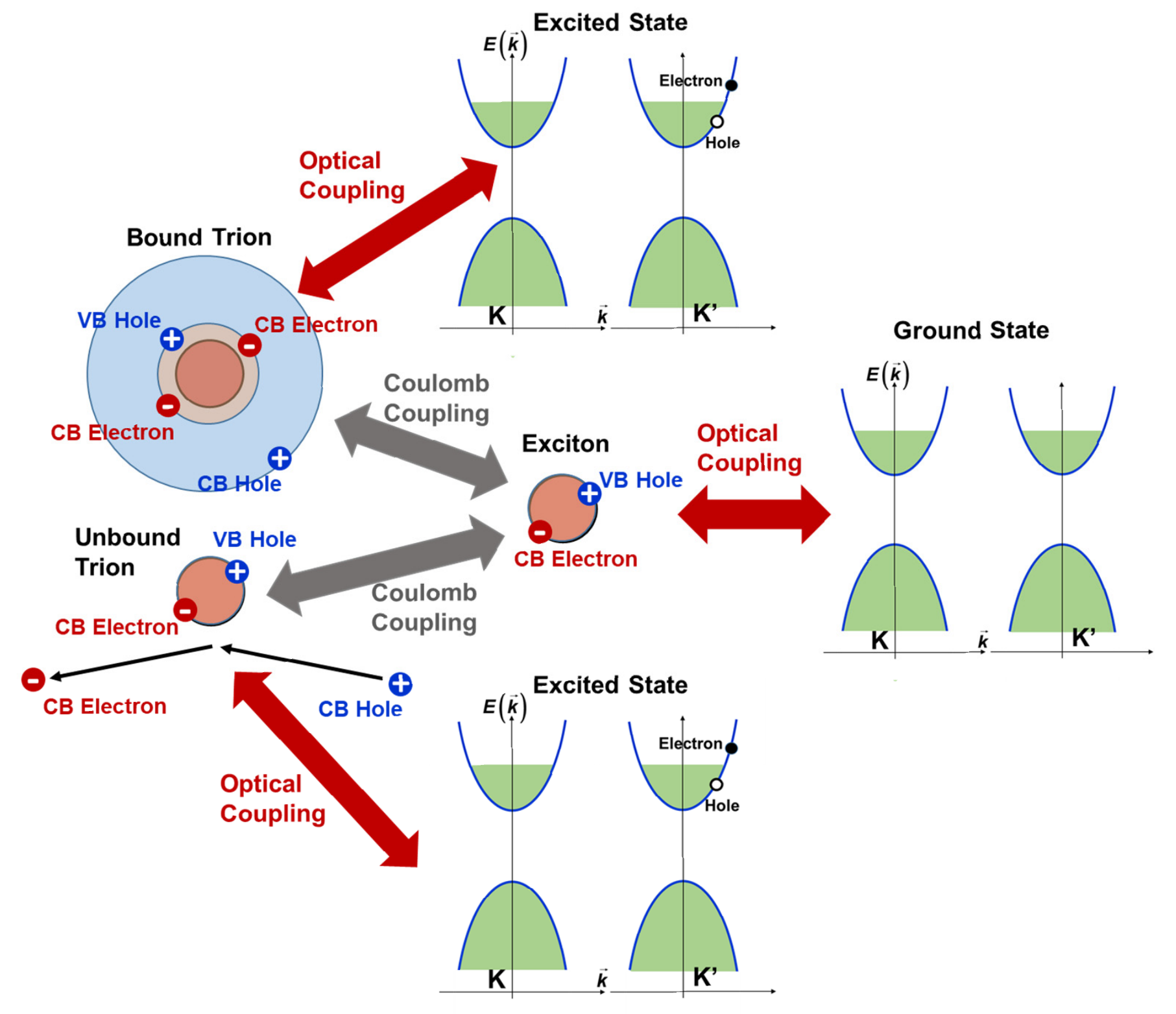}
   \caption{The nature of couplings involving 2-body exciton and 4-body trion states are depicted for an electron-doped material. The 4-body trion states are coupled to the 2-body exciton states via electron-electron and electron-hole Coulomb interactions. Only the exciton states are coupled to the material ground state via optical coupling. The trion states are optically coupled to excited states of the material consisting of a CB electron-hole pair. The trion states include both bound and unbound trion states.}
    \label{fig:fig1}
  \end{center}
\end{figure}

Several recent works have contributed to resolving this mystery and clarifying the nature of excitons and trions in doped semiconductors ~\cite{Rana20,Suris03,Macdonald17,Imam16,Chang19}. Recently, the authors have presented a theoretical model based on two coupled Schr{\"o}dinger equations to describe excitons and trions in electron-doped 2D materials~\cite{Rana20}. One is a 2-body Schr{\"o}dinger equation for a conduction band (CB) electron interacting with a valence band (VB) hole, and the other is a 4-body Schr{\"o}dinger equation of two CB electrons, one VB hole, and one CB hole interacting with each other. The CB hole is created when a CB electron is scattered out of the Fermi sea by an exciton. The eigenstates of the 2-body equation were identified with excitons and the eigenstates of the 4-body equation were identified with trions. A bound trion state is therefore a 4-body bosonic state, and not a 3-body fermionic state. The two Schr{\"o}dinger equations are coupled as a result of Coulomb interactions between the excitons and the trions in doped materials. The model shows that pure exciton and trion states are not eigenstates of the Hamiltonian in the presence of doping. However, good approximate eigenstates can be constructed from superpositions of exciton and trion states. This superposition includes both bound trion states as well as unbound trion states. The latter are exciton-electron scattering states. These superposition states, first proposed by Suris~\cite{Suris03}, resemble the exciton-polaron variational states proposed by Sidler et al.~\cite{Imam16,Macdonald17,Chang19}. The optical conductivity obtained from the model proposed by the authors explains all the prominent features experimentally seen in the optical absorption spectra of doped 2D materials including the observation of two prominent absorption peaks and the variation of their energy splittings and spectral shapes and strengths with the doping density~\cite{Rana20}. Furthermore, the peaks observed in the optical absorption spectra of doped 2D materials do not correspond to pure exciton or pure trion states. Each peak corresponds to a superposition of exciton and trion states.

While previous papers, including the one by the authors, have addressed the problem of light absorption by excitons and trions~\cite{Rana20,Suris03,Macdonald17,Chang19}, questions related to light emission and radiative lifetimes of excitons and trions in doped materials remain unanswered. The model developed by the authors~\cite{Rana20}, rather interestingly, also showed that the 4-body trion states have no optical matrix elements with the material ground state. The ground state of, say an electron-doped material, is defined as the state consisting of a completely full valence band (no VB holes), and a completely full Fermi sea in the conduction band (no CB holes inside and no CB electrons outside the Fermi sea). Therefore, the contribution to the material optical conductivity from the 4-body trion states results almost entirely from their Coulomb coupling to the 2-body exciton states~\cite{Rana20b}. The exciton and trion states and the related couplings are depicted in Fig.\ref{fig:fig1}. However, the trion states, including both bound and unbound trion states, are optically coupled to the excited states of the material consisting of a CB electron-hole pair. In other words, a trion state can decay by emitting a photon and leaving behind a CB electron-hole pair. The radiative rate of this process is significant after one has summed over all possible CB electron-hole pairs that can result from the radiative decay of a 4-body trion state. 

The experimentally relevant radiative lifetimes are not those of pure exciton and trion states, but of the approximate energy eigenstates which, as discussed above, are superpositions of exciton and trion states. The goal of this paper is to clarify the processes contributing to photon emission from these energy eigenstates in 2D materials and calculate the corresponding radiative lifetimes. Our main results are as follows. The radiative lifetimes of the exciton-trion energy eigenstates, with momenta inside the light cone, are found to be in the few hundred femtoseconds to a few picoseconds range and are strongly dependent on the doping density. Within the light cone, the exciton component of these eigenstates provides the dominant contribution to the radiative rates. The radiative lifetimes of the exciton-trion superposition states, with momenta outside the light cone, are in the few hundred picoseconds to a few nanoseconds range and are again strong functions of the doping density. Outside the light cone, only the trion component of these eigenstates contributes to the radiative rates. The doping density dependence of the radiative lifetimes of the two peaks in the optical emission spectra follows the doping density dependence of the spectral weights of the same two peaks observed in the optical absorption spectra as both have their origins in the Coulomb coupling between the excitons and trions in doped 2D materials.

\section{Theoretical Model} \label{sec:thmodel}
In this Section we set up the Hamiltonian and derive the main equations. Although the focus is on electron-doped 2D TMD materials, the arguments are kept general enough to be applicable to any 2D material.    

\subsection{The Hamiltonian} \label{subsec:hamiltonian}
\begin{figure}
  \begin{center}
   \includegraphics[width=0.7\columnwidth]{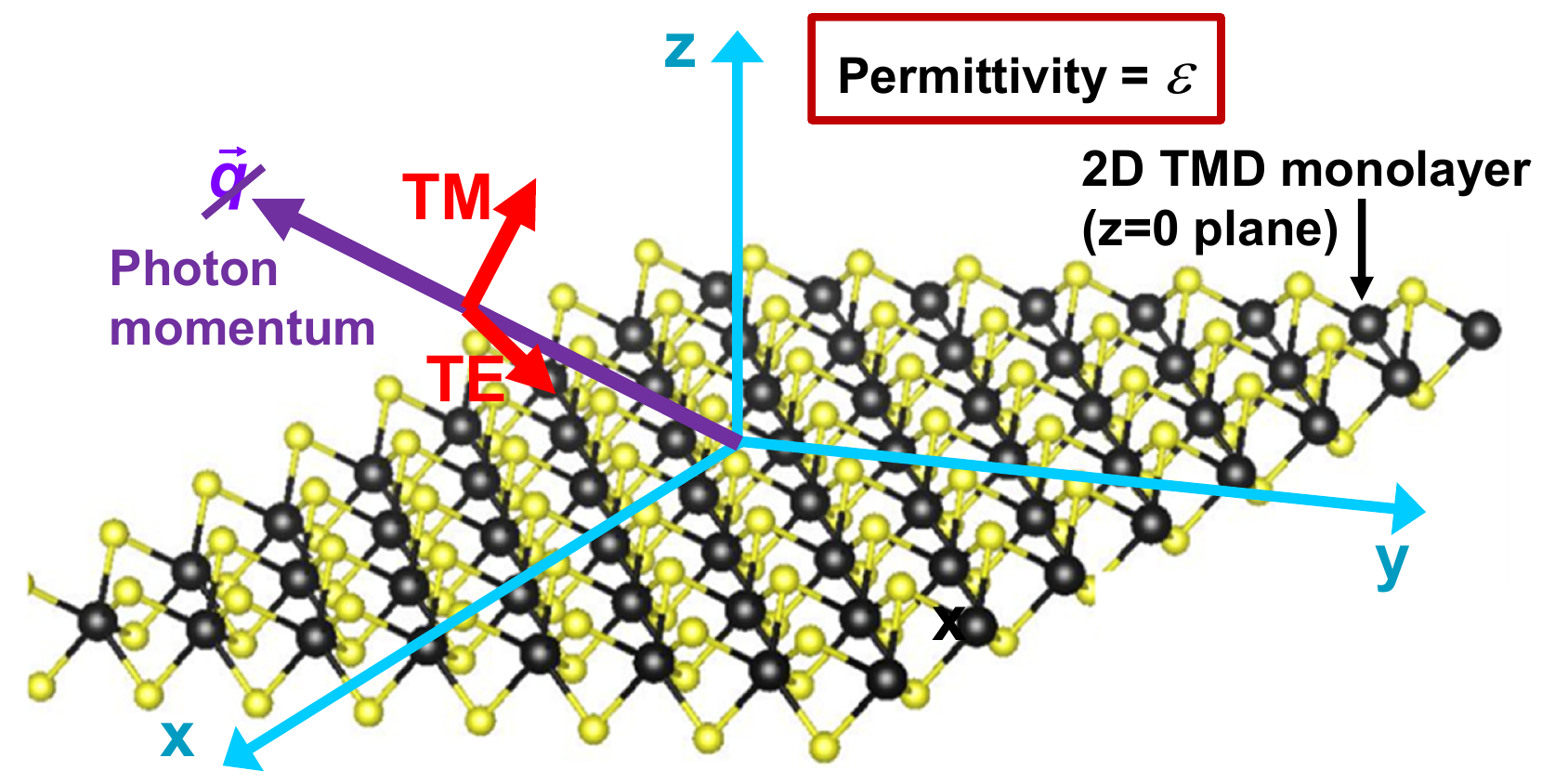}
    \caption{A 2D TMD monolayer in the $z=0$ plane is shown. The two light polarizations are also illustrated.}
    \label{fig:fig2}
  \end{center}
\end{figure}

We consider a 2D TMD monolayer located in the $z=0$ plane inside a uniform medium of dielectric constant $\epsilon$. The TMD layer interacts with both TE (electric field in the $z=0$ plane) and TM (magnetic field in the $z=0$ plane) polarized light modes. The Hamiltonian describing electrons and holes in the TMD layer (near the $K$ and $K'$ points in the Brillouin zone) interacting with each other and with the optical mode in the rotating wave approximation is~\cite{Xiao12,Changjian14,HWang16,Mano16},
\begin{eqnarray}
H & = & \sum_{\vec{k},s} E_{c,s}(\vec{k}) c_{s}^{\dagger}(\vec{k})c_{s}(\vec{k}) + \sum_{\vec{k},s} E_{v,s}(\vec{k}) b_{s}^{\dagger}(\vec{k})b_{s}(\vec{k}) \nonumber \\
& + & \frac{1}{A}\sum_{\vec{q},\vec{k},\vec{k}',s,s'} U(q)  c_{s}^{\dagger}(\vec{k}+\vec{q})b_{s'}^{\dagger}(\vec{k}'-\vec{q})b_{s'}(\vec{k}')c_{s}(\vec{k}) \nonumber \\
& + & \frac{1}{2A}\sum_{\vec{q},\vec{k},\vec{k}',s,s'} V(q)  c_{s}^{\dagger}(\vec{k}+\vec{q})c_{s'}^{\dagger}(\vec{k}'-\vec{q})c_{s'}(\vec{k}')c_{s}(\vec{k}) \nonumber \\
& + & \sum_{\vec{\slashed{q}},j} \hbar \omega(\slashed{q}) a^{\dagger}_{j}(\vec{\slashed{q}})a_{j}(\vec{\slashed{q}}) \nonumber \\
& + & \frac{1}{\sqrt{AL}}\sum_{q_{z},\vec{Q},\vec{k},j,s} \left( g_{j,s}(\vec{\slashed{q}}) c_{s}^{\dagger}(\vec{k}+\vec{Q})b_{s}(\vec{k})a_{j}(\vec{\slashed{q}})  + h.c \right) \nonumber \\
\label{eq:H}
\end{eqnarray}
Here, $E_{c,s}(\vec{k})$ and $E_{v,s}(\vec{k})$ are the conduction and valence band energies. $s,s'$ represent the spin/valley  degrees of freedom in the 2D material, and we assume for simplicity that the electron and hole effective masses are independent of the spin/valley. $U(\vec{q})$  represents Coulomb interaction between electrons in the conduction and valence bands and $V(\vec{q})$ represents Coulomb interaction among the electrons in the conduction bands. $A$ is the monolayer area and $AL$ is the volume assumed for field quantization. $\hbar \omega(\vec{\slashed{q}})$ is the energy of a photon with  momentum $\vec{\slashed{q}}$, and $g_{j,s}(\vec{\slashed{q}})$ is the electron-photon coupling constant for light with photon polarization $j = {{\rm TE},{\rm TM}}$ (see Fig.\ref{fig:fig2}). Most momentum vectors in the Hamiltonian above are in 2D. Those associated with light are in 3D, carry a slash in the notation for clarity, and $\vec{\slashed{q}} = \vec{Q} + q_{z}\hat{z}$, where $\vec{Q}$ is the momentum component in the $z=0$ plane. Other than for phase factors that are not relevant to the discussion in this paper, $g_{j,s}(\vec{\slashed{q}})$ for electron states near the band edges in 2D TMDs can be given by~\cite{HWang16,Mano16}, 
\begin{equation}
  g_{j,s}(\vec{\slashed{q}}) = ev\sqrt{\frac{\hbar}{2 \epsilon \omega(\vec{\slashed{q}})}} \times \left\{ \begin{array}{cc} q_{z}/\slashed{q} & \text{for TM} \\ 1 & \text{for TE} \end{array} \right.
\end{equation} 
where, $v$ is the interband velocity matrix element~\cite{Xiao12,Changjian14,HWang16,Mano16}. 

\subsection{Exciton States, Trion States, and Energy Eigenstates} \label{subsec:eigen}
As shown by Rana et al.~\cite{Rana20}, approximate eigenstates of the Hamiltonian in (\ref{eq:H}) can be written as a superposition of 2-body exciton and 4-body trion states,
\begin{eqnarray}
  & &  |\psi_{n,s}(\vec{Q})\rangle = \frac{\alpha_{n}}{\sqrt{A}} \sum_{k} \frac{\phi^{ex*}_{n,\vec{Q}}(\vec{k})}{N_{ex}}  \nonumber \\
  & & \times c^{\dagger}_{s}(\vec{k}+\lambda_{e}\vec{Q}) b_{s}(\vec{k}-\lambda_{h}\vec{Q}) |GS \rangle \nonumber \\
  && + \sum_{m,s'} \frac{\beta_{m,s'}}{\sqrt{A^{3}}} \sum^{\vec{\underline{k}}_{1},\vec{\underline{k}}_{2} \ne \vec{p}}_{\vec{k}_{1},\vec{k}_{2},\vec{p}} \frac{\phi^{tr*}_{m,\vec{Q}}(\vec{k}_{1},s;\vec{k}_{2},s';\vec{p},s')}{N_{tr}} \nonumber \\
  & & \times \, c^{\dagger}_{s}(\vec{\underline{k}}_{1}) c^{\dagger}_{s'}(\vec{\underline{k}}_{2}) b_{s}(\vec{\underline{k}}_{1} + \vec{\underline{k}}_{2}-(\vec{Q}+\vec{p})) c_{s'}(\vec{p}) |GS \rangle \nonumber \\
  \label{eq:var}
\end{eqnarray}
Here, $|GS \rangle$ is the ground state of the electron doped material. The normalization factors are,
\begin{eqnarray}
  & & N_{ex} = \sqrt{1-f_{c,s}(\vec{k}+\lambda_{e} \vec{Q})} \nonumber \\
  & & N_{tr} = \sqrt{(1 + \delta_{s,s'}) f_{c,s'}(\vec{p}) \left[ 1-f_{c,s}(\vec{\underline{k}}_{1}) \right]\left[ 1 - f_{c,s'}(\vec{\underline{k}}_{2}) \right]} \nonumber \\
\end{eqnarray}
The above energy eigenstate has (in-plane) momentum $\vec{Q}$. $\phi^{ex}_{n,\vec{Q}}(\vec{k}+\lambda_{h}\vec{Q})$ and $\phi^{tr}_{m,\vec{Q}}(\vec{k}_{1},s_{1};\vec{k}_{2},s_{2};\vec{p},s_{2})$ are eigenstates of the 2-body exciton and 4-body trion eigenequations, respectively~\cite{Rana20}. The corresponding eigenenergies are,  $E^{ex}_{n}(\vec{Q},s)$ and $E^{tr}_{m}(\vec{Q},s_{1},s_{2})$, respectively. $\lambda_{h} = 1-\lambda_{e} = m_{h}/m_{ex}$  ($m_{ex} = m_{e} + m_{h}$), where $m_{e}$ ($m_{h}$) is the electron (hole) effective mass. $m_{tr} = 2m_{e} + m_{h}$, $\xi = m_{e}/m_{tr}$, and $\eta = m_{h}/m_{tr}$. The underlined vector $\vec{\underline{k}}$ stands for $\vec{k}+\xi (\vec{Q}+\vec{p})$. The summation over the index $m$ implies summation over all bound and unbound trion states. Expressions for the coefficients $\alpha_{n}$ and $\beta_{m,s'}$ are given later in this paper. The states given above are good approximations to the actual eigenstates of  the Hamiltonian in (\ref{eq:H}) within the purview of single electron-hole pair excitations and provided one ignores multiple electron-hole pair excitations~\cite{Rana20}. In most cases of practical interest involving 2D TMDs, only the lowest energy exciton state needs to be considered. However, bound trion states as well as the continuum of unbound trion states need to be included since the energy differences involved therein are small~\cite{Rana20}. This makes the direct calculation of radiative rates using Fermi's Golden Rule awkward.

\begin{figure}
  \begin{center}
   \includegraphics[width=0.9\columnwidth]{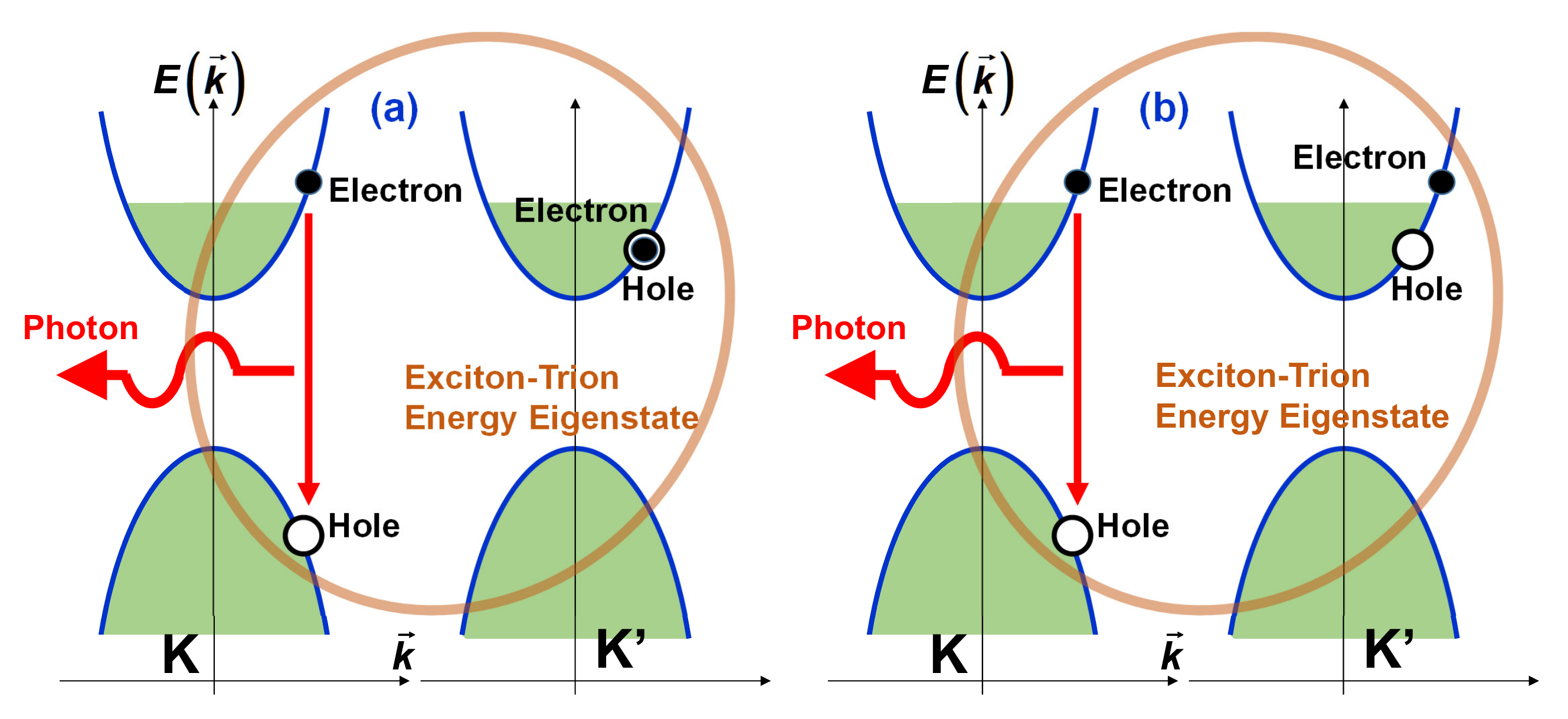}
   \caption{Two different kinds of photon emission processes are depicted. (a) Photon emission resulting in a decay of the energy eigenstate into the material ground state. The transition rate is determined by $|\alpha_{n}|^{2}$, the weight of the exciton component of the energy eigenstate in (\ref{eq:var}). This transition is possible only if the momentum $\vec{Q}$ of the energy eigenstate is within the light cone. (b) Photon emission resulting in a decay of the energy eigenstate into an excited state of the material that has a CB electron-hole pair. The CB electron-hole pair is left behind after photon emission from the trion components of the energy eigenstate. The transition rate is determined by $|\beta_{m,s'}|^{2}$ in (\ref{eq:var}). This transition is possible even if the momentum $\vec{Q}$ of the energy eigenstate is outside the light cone.}   
    \label{fig:fig6}
  \end{center}
\end{figure}

The optical interaction term in the Hamiltonian in (\ref{eq:H}) couples the material ground state to only the exciton component, and not to the trion components, in the exciton-trion supersposition energy eigenstates (see Fig.\ref{fig:fig1})~\cite{Rana20}. However, excited states of the material containing an electron-hole pair in the CB are optically coupled to the trion components. Given this, two different kinds of radiative transitions are possible and are depicted in Fig.\ref{fig:fig6}. Fig.\ref{fig:fig6}(a) shows photon emission resulting in a decay of the energy eigenstate into the material ground state. The transition rate is determined by $|\alpha_{n}|^{2}$, the weight of the exciton component of the energy eigenstate in (\ref{eq:var}). This transition is possible only if the momentum $\vec{Q}$ of the energy eigenstate is within the light cone. Fig.\ref{fig:fig6}(b) shows photon emission resulting in a decay of the energy eigenstate into an excited state of the material that has a CB electron-hole pair. The CB electron-hole pair is left behind after photon emission from the trion components of the energy eigenstate. Unlike the process in Fig.\ref{fig:fig6}(a), the process in Fig.\ref{fig:fig6}(b) is possible even if the momentum $\vec{Q}$ of the energy eigenstate is outside the light cone. If the emitted photon has an in-plane momentum $\vec{Q}'$ within the light cone, the difference $\vec{Q}-\vec{Q}'$ is taken by the electron-hole pair left behind in the CB. The radiative rate for this process is determined by the magnitude of the coefficients $\beta_{m,s'}$ of the trion states in the expression for the energy eigenstate given in (\ref{eq:var}).

In the Sections that follow, we will calculate separately the radiative rates for the two processes in Fig.\ref{fig:fig6}. 

\section{Rate for Radiative Decay into the Material Ground State} \label{sec:eom}
We first calculate the rate for the radiative decay of the energy eigenstate into the material ground state. This rate is expected to be proportional to the weight of the exciton component of the energy eigenstate, and the weight of the exciton component is conveniently given by the spectral density function which is proportional to the imaginary part of the exciton Green's function. Thus, we seek an expression for the radiative rate in terms of the exciton Green's function. 

\subsection{Heisenberg Equations}

We start from the Heisenberg equation for the photon operator,
\begin{equation}
  \left[ \hbar\omega(\vec{\slashed{q}}) + i\hbar \frac{d}{dt} \right] a^{\dagger}_{j}(\vec{\slashed{q}},t) = -\frac{1}{\sqrt{AL}} \sum_{\vec{k},s} g_{j,s}(\vec{\slashed{q}}) P_{\vec{Q}}(\vec{k},s;t)  \label{eq:photon1}
  \end{equation}
The polarization operator $P_{\vec{Q}}(\vec{k},s;t)$ equals $c_{s}^{\dagger}(\vec{k}+\vec{Q},t)b_{s}(\vec{k},t)$. The Heisenberg equation for the polarization operator is~\cite{Rana20},
\begin{eqnarray}
  & &  \left[ E_{c,s}(\vec{k}+\vec{Q}) - E_{v,s}(\vec{k}) + i\gamma_{ex} + i\hbar \frac{d}{dt} \right] P_{\vec{Q}}(\vec{k},s;t) = \nonumber \\
  & & - \frac{1}{\sqrt{AL}} \sum_{q_{z},j} g^{*}_{j,s}(\vec{\slashed{q}}) a^{\dagger}_{j}(\vec{\slashed{q}};t) \left[ 1 - f_{c,s}(\vec{k}+\vec{Q}) \right] + F_{\vec{Q}}(\vec{k},s;t) \nonumber \\
  & & + \frac{1}{A}\sum_{\vec{q}} U(\vec{q}) P_{\vec{Q}}(\vec{k} + \vec{q},s;t)  \left[ 1 - f_{c,s}(\vec{k} + \vec{Q}) \right] \nonumber \\
  & & -\frac{1}{A} \sum_{\vec{q},\vec{p},s'} U(\vec{q}) \nonumber \\
  & & \times T^{c}_{\vec{Q}}(\vec{k} + (\xi + \eta)\vec{Q} - \xi \vec{p},s;(\xi + \eta)\vec{p} - \xi \vec{Q} - \vec{q},s';\vec{p},s';t) \nonumber \\
  & & +\frac{1}{A} \sum_{\vec{q},\vec{p},s'} V(\vec{q}) \nonumber \\
  & & \times T^{c}_{\vec{Q}}(\vec{k} + (\xi + \eta)\vec{Q} - \xi \vec{p} +\vec{q},s;(\xi + \eta)\vec{p} - \xi \vec{Q} - \vec{q},s';\vec{p},s';t) \nonumber \\
  \label{eq:exciton1}
  \end{eqnarray}
Here, $ f_{c,s}(\vec{k})$ is the electron occupation probability in the conduction band (valence band is assumed to be completely full), $\gamma_{ex}$ is a phenomenological decoherence rate for the polarization that includes dephasing due to all processes other than exciton-electron scattering. $F_{\vec{Q}}(\vec{k},s;t)$ is a zero-mean delta-correlated quantum Langevin noise source that is introduced by the same processes that contribute to the decoherence $\gamma_{ex}$~\cite{haugbook}. The energies $E_{c,s}(\vec{k})$ include renormalizations due to exchange at the Hartree-Fock level ($-(1/A)\sum_{\vec{q}} V(\vec{q}) f_{c,s}(\vec{k}-\vec{q})$).  Taking the mean value of the operators in (\ref{eq:exciton1}), ignoring the first term and the last two terms on the right hand side (RHS), and Fourier transforming the remaining terms results in a 2-body Schr{\"o}dinger equation for the excitons~\cite{Rana20,haugbook,Kira12}. The last two terms in (\ref{eq:exciton1}) on the RHS contain four-body operators $T^{c}_{\vec{Q}}$. We define the operator $T_{\vec{Q}}(\vec{k}_{1},s_{1};\vec{k}_{2},s_{2};\vec{p},s_{2};t)$ as follows,
\begin{equation}
c_{s_{1}}^{\dagger}(\vec{\underline{k}}_{1};t) c_{s_{2}}^{\dagger}(\vec{\underline{k}}_{2};t) b_{s_{1}}(\vec{\underline{k}}_{1}+\vec{\underline{k}}_{2} - (\vec{Q}+\vec{p});t) c_{s_{2}}(\vec{p};t) 
\end{equation}
As before, the underlined vector $\vec{\underline{k}}$ stands for $\vec{k}+\xi (\vec{Q}+\vec{p})$. The average of the operator $T_{\vec{Q}}$ describes correlations arising from Coulomb interactions among four particles: two CB electrons, a VB hole, and a CB hole. $\vec{Q}$ is the total momentum of this 4-body state. We also define the connected operator $T^{c}_{\vec{Q}}$ as follows~\cite{Rana20},
\begin{eqnarray}
  & &  T_{\vec{Q}}(\vec{k}_{1},s_{1};\vec{k}_{2},s_{2};\vec{p},s_{2};t) = T^{c}_{\vec{Q}}(\vec{k}_{1},s_{1};\vec{k}_{2},s_{2};\vec{p},s_{2};t) \nonumber \\
  & & - f_{c,s_{2}}(\vec{p}) P_{\vec{Q}}(\vec{\underline{k}}_{1} -\vec{Q},s_{1};t) \delta_{\vec{\underline{k}}_{2},\vec{p}} \nonumber \\
  & & + f_{c,s_{2}}(\vec{p}) P_{\vec{Q}}(\vec{\underline{k}}_{2} - \vec{Q},s_{2};t) \delta_{s_{1},s_{2}} \delta_{\vec{\underline{k}}_{1},\vec{p}} \label{eq:conT}
\end{eqnarray}
The Heisenberg equation for the operator $T^{c}_{\vec{Q}}(\vec{k}_{1},s_{1};\vec{k}_{2},s_{2};\vec{p},s_{2})$ is found to be~\cite{Rana20},
\begin{eqnarray}
  & & \left[ E_{c,s_{1}}(\vec{\underline{k}}_{1}) + E_{c,s_{2}}(\vec{\underline{k}}_{2}) - E_{v,s_{1}}(\vec{\underline{k}}_{1}+\vec{\underline{k}}_{2} - (\vec{Q}+\vec{p})) \right. \nonumber \\
    & &  \left. - E_{c,s_{2}}(\vec{p}) + i\gamma_{tr} +i \hbar \frac{d}{dt} \right] T^{c}_{\vec{Q}}(\vec{k}_{1},s_{1};\vec{k}_{2},s_{2};\vec{p},s_{2};t) = \nonumber \\
  & & D_{\vec{Q}}(\vec{k}_{1},s_{1};\vec{k}_{2},s_{2};\vec{p},s_{2};t) \nonumber \\
  & & -\frac{1}{A} \sum_{\vec{q}} V(\vec{q}) T^{c}_{\vec{Q}}(\vec{k}_{1}+\vec{q},s_{1};\vec{k}_{2}-\vec{q},s_{2};\vec{p},s_{2};t) \nonumber \\
  & & \times \left[ 1-f_{c,s_{1}}(\vec{\underline{k}}_{1})  - f_{c,s_{2}}(\vec{\underline{k}}_{2}) \right] \nonumber \\
  & & +\frac{1}{A} \sum_{\vec{q}} U(\vec{q}) T^{c}_{\vec{Q}}(\vec{k}_{1}+\vec{q},s_{1};\vec{k}_{2},s_{2};\vec{p},s_{2};t) \left[ 1-f_{c,s_{1}}(\vec{\underline{k}}_{1})  \right] \nonumber \\
  & & +\frac{1}{A} \sum_{\vec{q}} U(\vec{q}) T^{c}_{\vec{Q}}(\vec{k}_{1},s_{1};\vec{k}_{2}-\vec{q},s_{2};\vec{p},s_{2};t) \left[ 1-f_{c,s_{2}}(\vec{\underline{k}}_{2}) \right] \nonumber \\
  & & +\frac{1}{A} \sum_{\vec{q}} V(\vec{q}) T^{c}_{\vec{Q}}(\vec{k}_{1}+ (\xi + \eta)\vec{q},s_{1};\vec{k}_{2}-\xi \vec{q},s_{2};\vec{p}+\vec{q},s_{2};t) \nonumber \\
  & & \times \left[f_{c,s_{2}}(\vec{p}) - f_{c,s_{1}}(\vec{\underline{k}}_{1}) \right] \nonumber \\
  & & +\frac{1}{A} \sum_{\vec{q}} V(\vec{q}) T^{c}_{\vec{Q}}(\vec{k}_{1} - \xi\vec{q},s_{1};\vec{k}_{2} + (\xi +\eta) \vec{q},s_{2};\vec{p}+\vec{q},s_{2};t) \nonumber \\
  & & \times \left[f_{c,s_{2}}(\vec{p}) - f_{c,s_{2}}(\vec{\underline{k}}_{2})  \right] \nonumber \\
  & & -\frac{1}{A} \sum_{\vec{q}} U(\vec{q}) T^{c}_{\vec{Q}}(\vec{k}_{1}-\xi \vec{q},s_{1};\vec{k}_{2}- \xi \vec{q},s_{2};\vec{p}+\vec{q},s_{2};t) f_{c,s_{2}}(\vec{p}) \nonumber \\
  & & + \frac{f_{c,s_{2}}(\vec{p})}{A} \sum_{\vec{q}} V(\vec{q}) \left[ 1-f_{c,s_{1}}(\vec{\underline{k}}_{1}) - f_{c,s_{2}}(\vec{\underline{k}}_{2})  \right] \nonumber \\
  & & \times \left[ P_{\vec{Q}}(\vec{\underline{k}}_{1} - \vec{Q} + \vec{q},s_{1};t) \delta_{\vec{\underline{k}}_{2}-\vec{q},\vec{p}} \right. \nonumber \\
  & & \left. - P_{\vec{Q}}(\vec{\underline{k}}_{2} - \vec{Q} - \vec{q},s_{2};t) \delta_{\vec{\underline{k}}_{1}+\vec{q},\vec{p}} \delta_{s_{1},s_{2}} \right] \nonumber \\
  & & - \frac{f_{c,s_{2}}(\vec{p})}{A} \sum_{\vec{q}} U(\vec{q}) \left\{ P_{\vec{Q}}(\vec{\underline{k}}_{1} - \vec{Q},s_{1};t) \delta_{\vec{\underline{k}}_{2}-\vec{q},\vec{p}} \left[ 1-f_{c,s_{2}}(\vec{\underline{k}}_{2})  \right] \right. \nonumber \\
  & & - \left. P_{\vec{Q}}(\vec{\underline{k}}_{2} - \vec{Q},s_{2};t) \delta_{\vec{\underline{k}}_{1}+\vec{q},\vec{p}} \delta_{s_{1},s_{2}} \left[ 1-f_{c,s_{1}}(\vec{\underline{k}}_{1})  \right] \right\} \nonumber \\
  \label{eq:4body1}
\end{eqnarray}
In deriving the above equation, all 6-body operator products were reduced to 4-body operator products using the random phase approximation~\cite{haugbook,Kira12}. By ignoring higher order correlations we are ignoring the generation of multiple particle-hole pairs in the CB. $\gamma_{tr}$ is a phenomenological decoherence rate and $D_{\vec{Q}}$ is the corresponding zero-mean delta-correlated Langevin noise source. If $\vec{r}_{e1}$, $\vec{r}_{e2}$, $\vec{r}_{h1}$, are $\vec{r}_{h2}$ the coordinates of the two electrons, the VB hole, and the CB hole, respectively, then $\vec{k}_{1}$, $\vec{k}_{2}$, $\vec{Q}$, and $\vec{p}$ are the momenta associated with the coordinates $\vec{r}_{e1} - \vec{r}_{h1}$, $\vec{r}_{e2} - \vec{r}_{h1}$, $\vec{R} = \xi (\vec{r}_{e1} + \vec{r}_{e2}) + \eta \vec{r}_{h1}$, and $\vec{R} - \vec{r}_{h2}$, respectively. Here, $\vec{R}$ is the center of mass coordinate of the two electrons and the VB hole. Taking the mean value of the operators in (\ref{eq:4body1}), ignoring the last two terms on the RHS in (\ref{eq:4body1}) that involve $P_{\vec{Q}}$, and Fourier transforming the remaining terms will result in a 4-body Schr{\"o}dinger equation for the trions~\cite{Rana20}. Each term on the RHS in the above equation (except the first and the last two) describes Coulomb interaction between two of the four particles. The last two terms involving $P_{\vec{Q}}$ describe the generation of four-body correlation from two-body correlations, or the creation of an CB electron-hole pair by an exciton.

We should mention here that a classical equation similar to (\ref{eq:4body1}) was obtained by Esser et al.~\cite{Esser01}. However, there are significant differences between (\ref{eq:4body1}) and the equation obtained by Esser et al.. In the work of Esser et al., the connected nature of $T^{c}_{\vec{Q}}$ was overlooked, the terms containing interactions with the CB hole were ignored, the phase-space restricting factors were ignored too, and, most importantly, the terms containing the polarization $P_{\vec{Q}}$ were also missed. Ignoring the coupling to $P_{\vec{Q}}$ in (\ref{eq:4body1}) is equivalent to ignoring exciton-trion coupling via Coulomb interactions. This coupling is responsible for making exciton-trion superposition states approximate eigenstates of the interacting system consisting of excitons and electrons in a doped material.

\subsection{Solution of Heisenberg Equations}
The polarization operator $P_{\vec{Q}}(\vec{k},s;t)$ can be decomposed using the complete set of exciton eigenfunctions~\cite{Rana20} $\phi^{ex}_{n,\vec{Q}}(\vec{k} + \lambda_{h}\vec{Q})$ as follows,
\begin{equation}
  P_{\vec{Q}}(\vec{k},s;t) =  \sum_{n} P_{n,\vec{Q}}(s;t) \sqrt{1-f_{c,s}(\vec{k} + \vec{Q})} \phi_{n,\vec{Q}}^{ex}(\vec{k} + \lambda_{h}\vec{Q}) \label{eq:decompose}
\end{equation}
We assume that at time $t$, $P_{n,\vec{Q}}(s;t)$ has a non-zero mean value for some particular values of $n$ and $s$. $\langle P_{n,\vec{Q}}(s;t) \rangle$ can be non-zero if the quantum state is a superposition of the material ground state $|GS\rangle$ and one of the eigenstates described in Section~\ref{subsec:eigen}. Following Milonni~\cite{Milonni}, the strategy going forward will then be as follows. The Heisenberg equations will be solved to find how the mean value $\langle P_{n,\vec{Q}}(s;t) \rangle$ decays with time due to radiative transitions, and the lifetime associated with this decay would give the radiative rate. Since we are exclusively interested in radiative transitions in this paper, several approximations will be made in order to keep the focus on the relevant physics and irrelevant terms will be ignored to keep the analysis simple. 

(\ref{eq:photon1}) can be be solved by direct integration to give,
\begin{eqnarray}
  & &  a^{\dagger}_{j}(\vec{\slashed{q}},t) = a^{\dagger}_{j}(\vec{\slashed{q}},t=0)e^{i\omega(\vec{\slashed{q}})t} \nonumber \\
  & & + \frac{i}{\sqrt{AL}} \sum_{\vec{k},s} \frac{g_{j,s}(\vec{\slashed{q}})}{\hbar} \int_{0}^{t} e^{i\omega(\vec{\slashed{q}})(t-t')} P_{\vec{Q}}(\vec{k},s;t') dt' \label{eq:photon2}
  \end{eqnarray}
Next, we find the time dependence of the operator $P_{n,\vec{Q}}(s;t)$. Using (\ref{eq:decompose}) in (\ref{eq:exciton1}), ignoring the Langevin noise sources on the RHS in (\ref{eq:exciton1}) and (\ref{eq:4body1}) (because these noise sources will not have any effect on the end results sought in this paper), and using the techniques discussed in a previous paper by the authors~\cite{Rana20} for solving the coupled system of equations in (\ref{eq:exciton1}) and (\ref{eq:4body1}), the operator $P_{n,\vec{Q}}(s;t)$ is found to be,
\begin{eqnarray}
  & & P_{n,\vec{Q}}(s;t) =  \int \frac{d\omega}{2\pi} \frac{-i\hbar e^{i\omega t} P_{n,\vec{Q}}(s;t=0)}{\hbar\omega - E^{ex}_{n}(\vec{Q},s) - i\gamma_{ex} - \Sigma^{ex*}_{n,s}(\vec{Q},\omega)}  \nonumber \\
  & & + \frac{1}{\sqrt{AL}} \sum_{q_{z},j} g^{*}_{j,s}(\vec{\slashed{q}}) \int \frac{d^{2}\vec{k}}{(2\pi)^{2}} \sqrt{1-f_{c,s}(\vec{k} + \vec{Q})} \phi_{n,\vec{Q}}^{ex*}(\vec{k} + \lambda_{h}\vec{Q}) \nonumber \\
  & & \times \int \frac{d\omega}{2\pi} \int_{0}^{t} \frac{e^{i\omega (t-t')} a^{\dagger}_{j}(\vec{\slashed{q}};t')}{\hbar\omega - E^{ex}_{n}(\vec{Q},s) - i\gamma_{ex} - \Sigma^{ex*}_{n,s}(\vec{Q},\omega)} \nonumber \\
  \label{eq:pol1}
\end{eqnarray}
Here, $\Sigma^{ex}_{n,s}(\vec{Q},\omega)$ is the self-energy of the excitons arising from their Coulomb coupling to the trions~\cite{Rana20},
\begin{eqnarray}
  &&  \Sigma^{ex}_{n,s}(\vec{Q},\omega) = \sum_{m,s'} \frac{ (1 + \delta_{s,s'}) \left| M_{m,n}(\vec{Q},s,s') \right|^{2}}{\hbar\omega - E^{tr}_{m}(\vec{Q},s,s') + i\gamma_{tr} }  \nonumber \\
  \label{eq:self1}
\end{eqnarray}
The summation over $m$ above implies a summation over all bound and unbound trion states consistent with the values of $s$ and $s'$. The expression for the Coulomb matrix elements $M_{m,n}(\vec{Q},s,s')$ coupling the exciton and trion states can be found in a previous paper by Rana et al.~\cite{Rana20}. The exciton self-energy thus includes contribution of trion states to the polarization via exciton-trion Coulomb coupling. (\ref{eq:pol1}) gives the natural frequencies associated with the material polarization response, given by the poles of the expression in the denominator, and these frequencies also correspond to the energy eigenstates of the Hamiltonian~\cite{Rana20}. It follows that on fast time scales (of the order of the inverse of the relevant optical frequencies), $P_{n,\vec{Q}}(s;t)$ can be written as,
\begin{eqnarray}
  & &  P_{n,\vec{Q}}(s;t') \approx P_{n,\vec{Q}}(s;t) \times \nonumber \\
  & & \left\{ \begin{array}{lr}
  {\displaystyle \int \frac{d\omega}{2\pi} \frac{-i\hbar e^{-i\omega (t-t')}}{\hbar\omega - E^{ex}_{n}(\vec{Q},s) - i\gamma_{ex} - \Sigma^{ex*}_{n,s}(\vec{Q},\omega)}} & t'>t  \\
  {\displaystyle \int \frac{d\omega}{2\pi} \frac{i\hbar e^{-i\omega (t-t')}}{\hbar\omega - E^{ex}_{n}(\vec{Q},s) + i\gamma_{ex} - \Sigma^{ex}_{n,s}(\vec{Q},\omega)}} & t'<t \end{array} \right. \nonumber \\
\end{eqnarray}
The above approximation, when used together with (\ref{eq:decompose}) in (\ref{eq:photon2}), results in an expression for the photon operator in the standard Markoff approximation~\cite{Milonni},
\begin{eqnarray}
  & &  a^{\dagger}_{j}(\vec{\slashed{q}},t) = a^{\dagger}_{j}(\vec{\slashed{q}},t=0)e^{i\omega(\vec{\slashed{q}})t} \nonumber \\
  & & - \sqrt{\frac{A}{L}} \sum_{n,s} g_{j,s}(\vec{\slashed{q}})   \int \frac{d^{2}\vec{k}}{(2\pi)^{2}} \sqrt{1-f_{c,s}(\vec{k} + \vec{Q})} \phi_{n,\vec{Q}}^{ex}(\vec{k} + \lambda_{h}\vec{Q}) \nonumber \\
  & & \times \frac{P_{n,\vec{Q}}(s;t)}{\hbar\omega(\vec{\slashed{q}}) - E^{ex}_{n}(\vec{Q},s) + i\gamma_{ex} - \Sigma^{ex}_{n,s}(\vec{Q},\omega)}  \nonumber \\
  \label{eq:photon3}
  \end{eqnarray}  

\subsection{Radiative Rate}
Use of (\ref{eq:photon3}) in the first term on the RHS of (\ref{eq:exciton1}) introduces an additional source of damping in the material polarization which is due to radiative transitions. To show this more clearly, we substitute (\ref{eq:photon3}) in (\ref{eq:exciton1}), then use the decomposition in (\ref{eq:decompose}) and project out the equation for $P_{n,\vec{Q}}(s;t)$, take the mean value, and retain only those terms that are relevant to see this radiative damping to get,
\begin{equation}
 \frac{d\langle P_{n,\vec{Q}}(s;t)\rangle}{dt} \sim - \frac{R_{n,s}(\vec{Q})}{2} \langle P_{n,\vec{Q}}(s;t)\rangle
  \label{eq:photon4}
  \end{equation}  
where the spontaneous emission rate $R_{n,s}(\vec{Q})$ is,
\begin{eqnarray}
  & & R_{n,s}(\vec{Q}) = \frac{2}{c\epsilon}\int^{\infty}_{Qc} \frac{d\omega}{2\pi} \left( \frac{\omega}{\sqrt{\omega^{2} - Q^2c^2}} + \frac{\sqrt{\omega^{2} - Q^2c^2}}{\omega} \right)  \nonumber \\
  & & \times \text{Re} \left[ \sigma_{n,s}(\vec{Q},\omega) \right] \label{eq:rate1}
\end{eqnarray}
Here, $c=1/\sqrt{\epsilon \mu_{o}}$ is the speed of light in the medium surrounding the 2D monolayer. The above result for the spontaneous emission is conveniently expressed in terms of the relevant exciton/trion optical conductivity of the 2D TMD monolayer. (\ref{eq:rate1}) is the main result of this paper. The optical conductivity of a 2D TMD monolayer, for in-plane light polarization, can be written in terms of the exciton Green's function~\cite{Rana20},
\begin{eqnarray}
  &&  \sigma(\vec{Q},\omega) = \sum_{n,s} \sigma_{n,s}(\vec{Q},\omega) \nonumber \\
  && = i\frac{e^{2}v^{2}}{\omega} \sum_{n,s} \left|\int \frac{d^{2}\vec{k}}{(2\pi)^{2}} \phi^{ex}_{n,\vec{Q}}(\vec{k} + \lambda_{h}\vec{Q}) \sqrt{1-f_{c,s}(\vec{k} + \vec{Q})} \right|^{2} \nonumber \\
  & & \times G^{ex}_{n,s}(\vec{Q},\omega) \nonumber \\
  \label{eq:cond1}
\end{eqnarray}
Here, $G^{ex}_{n,s}(\vec{Q},\omega)$ is the exciton Green's function~\cite{Rana20},
\begin{equation}
  G^{ex}_{n,s}(\vec{Q},\omega) = \frac{1}{\hbar\omega - E^{ex}_{n}(\vec{Q},s) + i\gamma_{ex} - \Sigma^{ex}_{n,s}(\vec{Q},\omega)} \label{eq:green}
\end{equation}
The energies of the eigenstates in (\ref{eq:var}) are given by the poles of the exciton Green's function. We label these energies as $E^{lo}_{n,s}(\vec{Q})$ and $E^{hi}_{n,s}(\vec{Q})$. Earlier, in Section~\ref{subsec:eigen}, we had remarked that the radiative rate for the energy eigenstate to decay into the ground state is proportional to the weight of its exciton component given by $\alpha_{n}$  in (\ref{eq:var}). Assuming, $\gamma_{tr}=\gamma_{ex}=0$ for simplicity, $|\alpha_{n}|^{2}$ for an energy eigenstate equals the residue of the exciton Green's function at the energy of the eigenstate,
\begin{eqnarray}
 |\alpha_{n}|^{2} & = & \left[ 1 - \frac{1}{\hbar}\frac{\partial }{\partial \omega}\text{Re} \Sigma^{ex}_{n,s}(\vec{Q},\omega) \right]^{-1} \nonumber \\
  & = & \frac{1}{\displaystyle 1 + \sum_{m,s'} \frac{\displaystyle (1 + \delta_{s,s'}) \left| M_{m,n}(\vec{Q},s,s') \right|^{2}}{\displaystyle \left(E^{lo/hi}_{n,s}(\vec{Q}) - E^{tr}_{m}(\vec{Q},s,s') \right)^{2} }}  \nonumber \\
\end{eqnarray}

\begin{figure}
  \begin{center}
   \includegraphics[width=0.8\columnwidth]{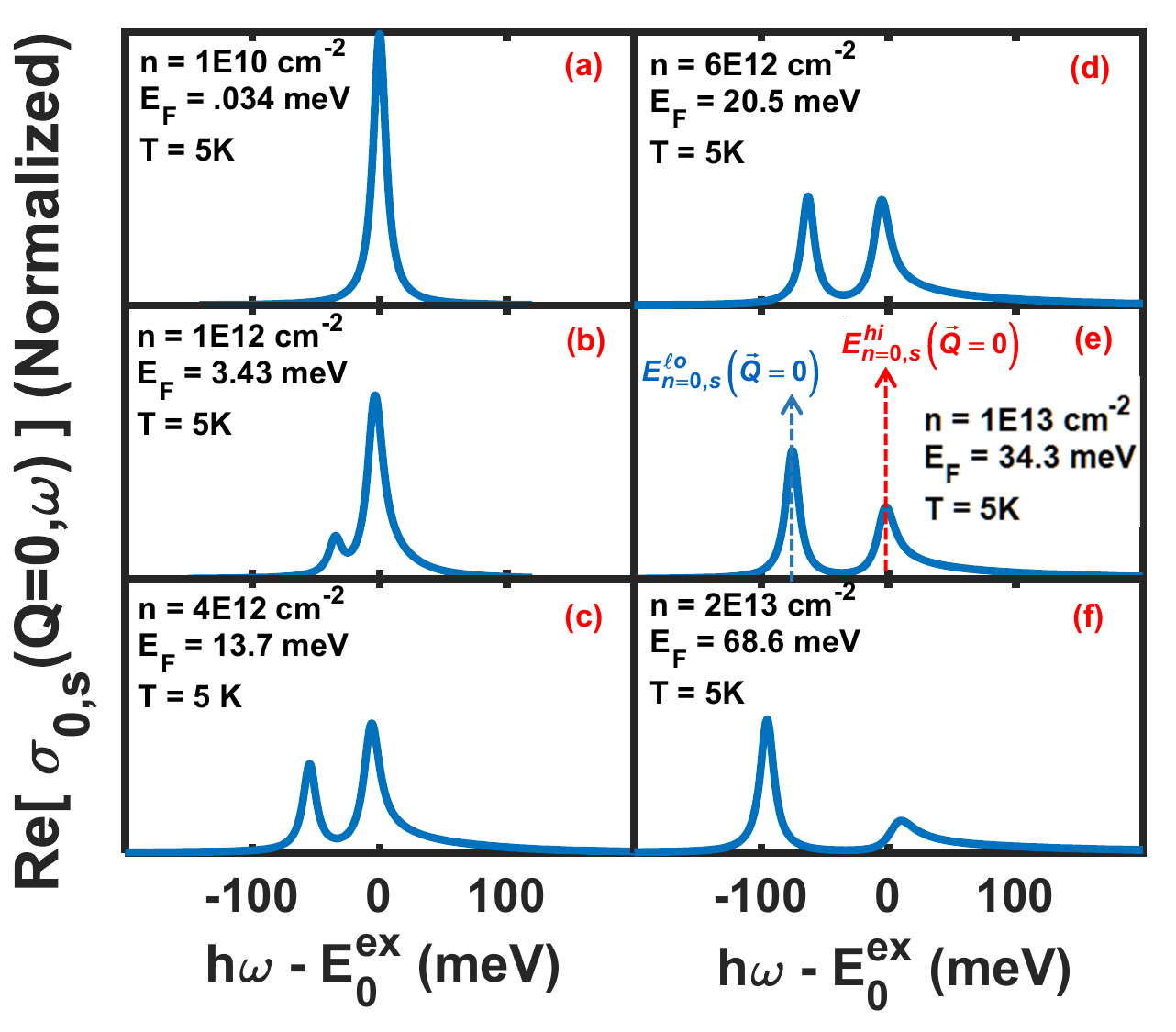}
   \caption{Calculated real part of the optical conductivity, $\sigma_{0,s}(\vec{Q}=0,\omega)$, for in-plane light polarization is plotted for different electron densities for electron-doped monolayer 2D MoSe$_{2}$. Only the lowest energy exciton state is considered in the calculations. The spectra are all normalized to the peak optical conductivity value at zero electron density. T = 5K. The frequency axis is offset by the exciton eigenenergy $E^{ex}_{0}(\vec{Q}=0,s)$ of the two-body Schr{\"o}dinger equation. Two prominent peaks are seen in the spectra. Each peak corresponds to an energy eigenstate state that is a superposition of exciton and trion states, as shown in (\ref{eq:var}). Figure is reproduced from the paper by Rana et al.~\cite{Rana20}.}   
    \label{fig:fig3}
  \end{center}
\end{figure}

Before exploring the above results further, it is instructive look at the optical conductivity of 2D materials. The exciton/trion optical conductivity of electron-doped 2D MoSe$_{2}$ was calculated by the authors in a recent paper and the results are reproduced in Fig.\ref{fig:fig3}~\cite{Rana20}. The spectra shows two prominent absorption peaks which correspond to the poles, $E^{lo}_{n,s}(\vec{Q})$ and $E^{hi}_{n,s}(\vec{Q})$, of the exciton Green's function in (\ref{eq:green}). The spectral weight shifts from the higher energy peak to the lower energy peak as the electron density increases. The energy separation between the two peaks also increases nearly linearly with the electron density~\cite{Rana20}. In the literature, the lower energy absorption peak is often identified with the trions (or charged excitons) and the higher energy peak with the excitons. This identification is true only in the limit of very small electron densities. At electron densities large enough such that the lower energy peak has sufficient spectral weight to be experimentally visible in the absorption spectrum, each peak corresponds to an energy eigenstate that is a superposition of exciton and trion states, as shown in (\ref{eq:var}). Furthermore, at large electron densities, the higher energy peak is broadened due to exciton-electron scattering and acquires a wide pedestal (more visible on its higher energy side) that corresponds to the continuum of unbound trion states (or exciton-electron scattering states). In Fig.\ref{fig:fig3}, linewidth broadening due to factors other than exciton-electron scattering, such as phonon scattering, was included by assuming that $\gamma_{ex}=\gamma_{tr}=4$ meV.    

The rates, $R^{lo}_{n,s}(\vec{Q})=1/\tau^{lo}_{n,s}(\vec{Q})$ and $R^{hi}_{n,s}(\vec{Q})=1/\tau^{hi}_{n,s}(\vec{Q})$, corresponding to the lower and higher energy peaks in the absorption spectra, respectively, can be each obtained by restricting the frequency integral in (\ref{eq:rate1}) to the respective peak. Interestingly, because the integral of the optical conductivity in (\ref{eq:cond1}) satisfies the sum rule~\cite{Rana20},
\begin{equation}
  \int_{0}^{\infty} \omega {\rm Re}\{\sigma(\vec{Q},\omega)\} \, \frac{d\omega}{2\pi} = \frac{e^{2}v^{2}}{2\hbar} \sum_{s} \int \frac{\displaystyle d^{2}\vec{k}}{\displaystyle (2\pi)^{2}} \left( 1 -  f_{c,s}(\vec{k}) \right) \label{eq:sum}
\end{equation}
one can expect from (\ref{eq:rate1}) that the radiative rate for the lower energy absorption peak to increase with the electron density and the radiative rate for the higher energy absorption peak to decrease with the electron density such that the sum rule above is always satisfied. In addition, since the area under the two peaks in Fig.\ref{fig:fig3} become nearly the same at large electron densities (~$2\times 10^{13}$ cm$^{-2}$) (despite the fact that the peak optical conductivity of the lower energy peak is higher), one can expect the two lifetimes to become comparable at large electron densities. Numerical simulation results, presented in the next Section, confirm these findings. 

\begin{figure}
  \begin{center}
   \includegraphics[width=0.9\columnwidth]{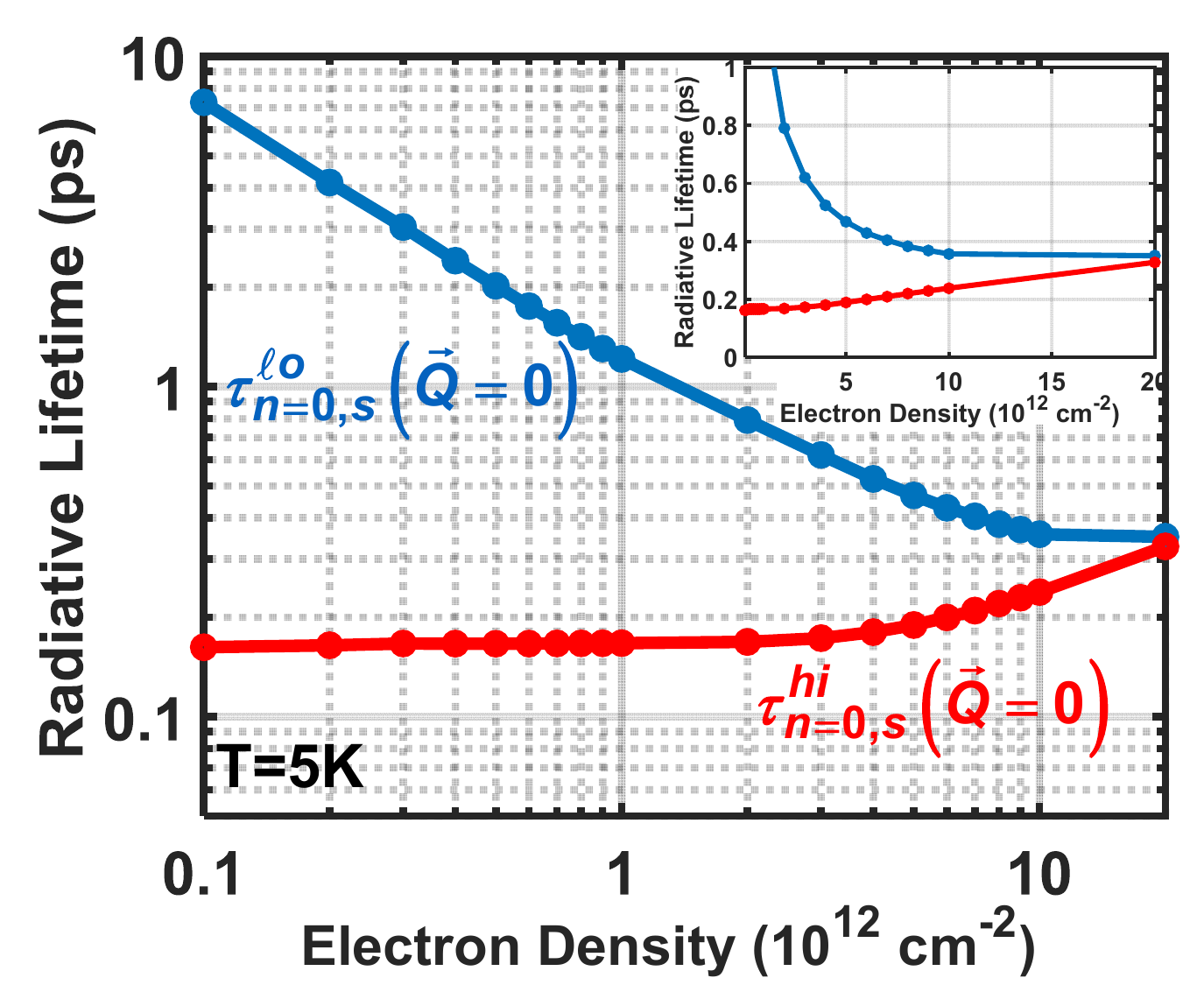}
   \caption{The zero-momentum radiative lifetimes,$\tau^{lo}_{n=0,s}(\vec{Q}=0)$ and $\tau^{hi}_{n-0,s}(\vec{Q}=0)$, of the lower and higher energy eigenstates, respectively, of the coupled exciton-trion system (and corresponding to the lower and higher energy peaks in the optical absorption spectra in Fig.\ref{fig:fig3}) are plotted as a function of the electron densities for an electron-doped monolayer 2D MoSe$_{2}$ suspended in air. T=5 K. The inset shows the same data on a linear scale.}   
    \label{fig:fig4}
  \end{center}
\end{figure}

\begin{figure}
  \begin{center}
   \includegraphics[width=0.9\columnwidth]{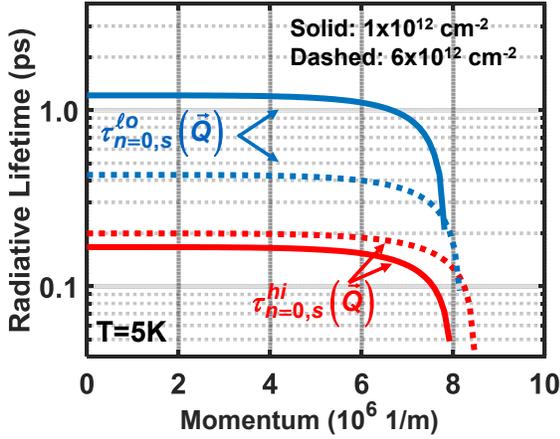}
   \caption{The radiative lifetimes,$\tau^{lo}_{n=0,s}(\vec{Q})$ and $\tau^{hi}_{n-0,s}(\vec{Q})$, of the lower and higher energy eigenstates, respectively, of the coupled exciton-trion system (and corresponding to the lower and higher energy peaks in the optical absorption spectra in Fig.\ref{fig:fig3}) are plotted as a function of the in-plane momentum $Q$ for different electron densities ($10^{12}$ cm$^{-2}$ and $6\times10^{12}$ cm$^{-2}$) for an electron-doped monolayer 2D MoSe$_{2}$ suspended in air. T=5 K.}   
    \label{fig:fig5}
  \end{center}
\end{figure}

\subsection{Numerical Simulations and Results}
For simulations, we consider an electron-doped monolayer of 2D MoSe$_{2}$ suspended in air. In monolayer MoSe$_{2}$, spin-splitting of the conduction bands is large ($\sim$35 meV~\cite{Kosmider13}) and the lowest conduction band in each of the $K$ and $K'$ valleys is optically coupled to the topmost valence band~\cite{Xiao13}. We use effective mass values of $0.7 m_{o}$ for both $m_{e}$ and $m_{h}$ which agree with the recently measured value of $0.35 m_{o}$ for the exciton reduced mass~\cite{Goryca19}. We use a wavevector-dependent dielectric constant $\epsilon(\vec{q})$, appropriate for 2D materials~\cite{Changjian14}, to screen the Coulomb potentials. We assume that $\gamma_{ex}=\gamma_{tr}\sim 4$ meV~\cite{Knorr16}. We compute exciton and trion eigenfunctions and eigenenergies for different momenta and electron densities as described by Rana et al.~\cite{Rana20}. 

Fig.\ref{fig:fig4} shows the zero-momentum radiative lifetimes, $\tau^{lo}_{n=0,s}(\vec{Q}=0)$ and $\tau^{hi}_{n-0,s}(\vec{Q}=0)$, plotted for different electron densities. As expected, at very small electron densities the radiative lifetime $\tau^{lo}_{n=0,s}(\vec{Q}=0)$ of the lower energy eigenstate is much longer than the lifetime $\tau^{hi}_{n=0,s}(\vec{Q}=0)$ of the higher energy eigenstate. At very large electron densities these two lifetimes become comparable. At small electron densities, when the entire spectral weight lies with the higher energy absorption peak in Fig.\ref{fig:fig3}, and the corresponding eigenstate is essentially a pure exciton state, the calculated lifetimes for the higher energy eigenstate agree well with the lifetimes published previously for excitons in 2D materials~\cite{HWang16,Grossman15}. But at larger electron densities (>$10^{12}$ 1/cm$^{2}$), the results in previous work, which treated excitons and trions as independent excitations, become incorrect.        

Fig.\ref{fig:fig5} shows the radiative lifetimes,$\tau^{lo}_{n=0,s}(\vec{Q})$ and $\tau^{hi}_{n-0,s}(\vec{Q})$, plotted as a function of the in-plane momentum $Q$ (within the light cone) for different electron densities. The light cone momentum is defined as the momentum $Q$ for which the energy of the eigenstate, $E^{lo}_{n,s}(\vec{Q})$ or $E^{hi}_{n,s}(\vec{Q})$, equals the photon energy $\hbar Q c$. The radiative lifetimes are more or less constant for momenta within the light cone, decrease rapidly as the momentum approaches the light cone (due to an increase in the density of photon states), and then diverge for momenta outside the light cone (where the excitonic component of the energy eigenstates cannot emit a photon and decay into the material ground state). This behavior is well known for pure exciton states in 2D materials~\cite{HWang16,Grossman15,haugbook}, and it carries over to the coupled exciton-trion energy eigenstates in doped 2D materials.

\section{Rate for Radiative Decay into the Material Excited States} 
The radiative rates calculated above correspond to the process depicted in Fig.\ref{fig:fig6}(a) in which the energy eigenstate decays into the material ground state. In this Section, we calculate the radiative rate for the process in Fig.\ref{fig:fig6}(b) in which the energy eigenstate decays into an excited state of the material that has an electron-hole pair in the CB. The final state after photon emission consists of a photon with momentum $\vec{q'}=\hat{z}q'_{z} + \vec{Q'}$, a CB hole with momentum $\vec{p}$ and a CB electron with momentum $\vec{p}+\vec{Q}-\vec{Q'}$. The radiative rate expression must include a summation over all these final states. Furthermore, the radiative rate for the process in Fig.\ref{fig:fig6}(b) is expected to be determined by the magnitude of the coefficients $\beta_{m,s'}$ of the trion states in the expression for the energy eigenstate given in (\ref{eq:var}). These coefficients are found to be,
  \begin{equation}
    |\beta_{m,s'}|^{2}  =  \frac{ \frac{\displaystyle (1 + \delta_{s,s'}) \left| M_{m,n}(\vec{Q},s,s') \right|^{2}}{\displaystyle \left(E^{lo/hi}_{n,s}(\vec{Q}) - E^{tr}_{m}(\vec{Q},s,s') \right)^{2}}}{\displaystyle 1 + \sum_{m',s''} \frac{\displaystyle (1 + \delta_{s,s''}) \left| M_{m',n}(\vec{Q},s,s'') \right|^{2}}{\displaystyle \left(E^{lo/hi}_{n,s}(\vec{Q}) - E^{tr}_{m'}(\vec{Q},s,s'') \right)^{2} }}  \nonumber \\ \label{eq:beta}
  \end{equation}
The summation over $m'$ above implies a summation over all bound and unbound trion states consistent with the values of $s$ and $s''$. The expression for the Coulomb matrix elements $M_{m,n}(\vec{Q},s,s')$ coupling the exciton and trion states (including bound and unbound trion states) can be found in a previous paper by Rana et al.~\cite{Rana20}.

\subsection{Radiative Rate}
  In order to calculate the radiative rates for the process in Fig.\ref{fig:fig6}(b), we avoid truncating the 6-body operator products to 4-body operator products that appear during the derivation of (\ref{eq:4body1}), and then include a Heisenberg equation for 6-body operator products in our model. The calculations are tedious and not particularly illuminating. The final result for the radiative rate $R_{n,s}(\vec{Q})$ can be written in a simple form,
\begin{eqnarray}
  & & R_{n,s}(\vec{Q}) = \sum_{m,s'} \frac{e^{2}v^{2}}{\epsilon} (1 + \delta_{s,s'}) \int \frac{dq'_{z}}{2\pi} \int \frac{d^{2}\vec{Q'}}{(2\pi)^{2}} \int \frac{d^{2}\vec{p}}{(2\pi)^{2}} \nonumber \\
  & & \times \left[ 1+ \frac{{q'}_{z}^{2}}{{Q'}^{2} + {q'}_{z}^{2}} \right] \left| \int \frac{d^{2}\vec{k}}{(2\pi)^{2}}  \right. \nonumber \\
  & &  \times \phi^{tr}_{m,\vec{Q}}(\vec{k} - \xi(\vec{Q} + \vec{p}), s; (\xi + \eta)(\vec{Q} + \vec{p}) - \vec{Q'},s';\vec{p'},s')    \nonumber \\
  & & \left. \times \sqrt{1 - f_{c,s}(\vec{k})} \right|^{2} \text{Re}\left[ \frac{i}{\omega} S_{n,s,m,s'}(\vec{Q},\vec{p},\vec{Q'},\omega)| \right]_{\omega = \sqrt{{q'}^{2}_{z}+{Q'}^{2}}c} \nonumber \\ \label{eq:rate2}
\end{eqnarray}
The spectral function $S_{n,s,m,s'}(\vec{Q},\vec{p},\vec{Q'},\omega)$ is,
\begin{eqnarray}
  & &   S_{n,s,m,s'}(\vec{Q},\vec{p},\vec{Q'},\omega) = \nonumber \\
  & & \frac{1}{\hbar \omega - E^{tr}_{m}(\vec{Q},s,s') + \Delta + i\gamma_{tr} - \Sigma_{n,s,m,s'}(\vec{Q},\vec{p},\vec{Q'},\omega)} \nonumber \\
\end{eqnarray}
Here, $\Delta$ stands for the energy difference $E_{c,s'}(\vec{p} + \vec{Q} - \vec{Q'}) - E_{c,s'}(\vec{p})$, and, 
\begin{eqnarray}
& & \Sigma_{n,s,m,s'}(\vec{Q},\vec{p},\vec{Q'},\omega) = \nonumber \\
& & \frac{(1 + \delta_{s,s'}) \left| M_{m,n}(\vec{Q},s,s') \right|^{2}}{\hbar\omega - E^{ex}_{n}(\vec{Q},s) + \Delta + i\gamma_{tr}   - F_{n,s,m,s'}(\vec{Q},\vec{p},\vec{Q'},\omega) } \nonumber \\
\end{eqnarray}
where,
\begin{eqnarray}
  & &   F_{n,s,m,s'}(\vec{Q},\vec{p},\vec{Q'},\omega) = \nonumber \\
  & & \sum_{m'\ne m, s''\ne s'} \frac{(1 + \delta_{s,s''}) \left| M_{m',n}(\vec{Q},s,s'') \right|^{2}}{ \hbar \omega - E^{tr}_{m}(\vec{Q},s,s'') + \Delta + i\gamma_{tr}} \nonumber \\
  \end{eqnarray}
The spectral function $S_{n,s,m,s'}(\vec{Q},\vec{p},\vec{Q'},\omega)$ has the following two important properties:
\begin{itemize}
\item Its poles are at the energies of the exciton-trion superposition eigenstates shifted by $\Delta$, the energy taken by the electron-hole pair left behind in the CB after photon emission. Therefore, the spectrum of $S_{n,s,m,s'}(\vec{Q},\vec{p},\vec{Q'},\omega)$ will have two prominent peaks just like the spectrum of optical absorption. Since for $Q<<k_{F}$, the energy shift $\Delta$ will be negligibly small for all $p<k_{F}$, and the peaks in the $S_{n,s,m,s'}(\vec{Q},\vec{p},\vec{Q'},\omega)$ spectrum will be more or less at the same energies as the peaks in the absorption spectrum. 
\item Assuming $\gamma_{ex}=\gamma_{tr}=0$, the residue of $S_{n,s,m,s'}(\vec{Q},\vec{p},\vec{Q'},\omega)$ at these two poles is exactly equal to the values of $|\beta_{m,s'}|^{2}$ given in (\ref{eq:beta}), which is satisfying in the light of the discussion above. 
\end{itemize}
The radiative rates, $R^{lo}_{n,s}(\vec{Q})=1/\tau^{lo}_{n,s}(\vec{Q})$ and $R^{hi}_{n,s}(\vec{Q})=1/\tau^{hi}_{n,s}(\vec{Q})$, corresponding to the lower and higher energy peaks in the absorption spectra, respectively, and associated with the process shown in Fig.\ref{fig:fig6}(b), can be each obtained by restricting the frequency integral in (\ref{eq:rate2}) to the respective spectral peak (the integral over frequency is implicit in (\ref{eq:rate2}) in the $q'_{z}$ and $\vec{Q'}$ integrations).     

\begin{figure}
  \begin{center}
   \includegraphics[width=0.9\columnwidth]{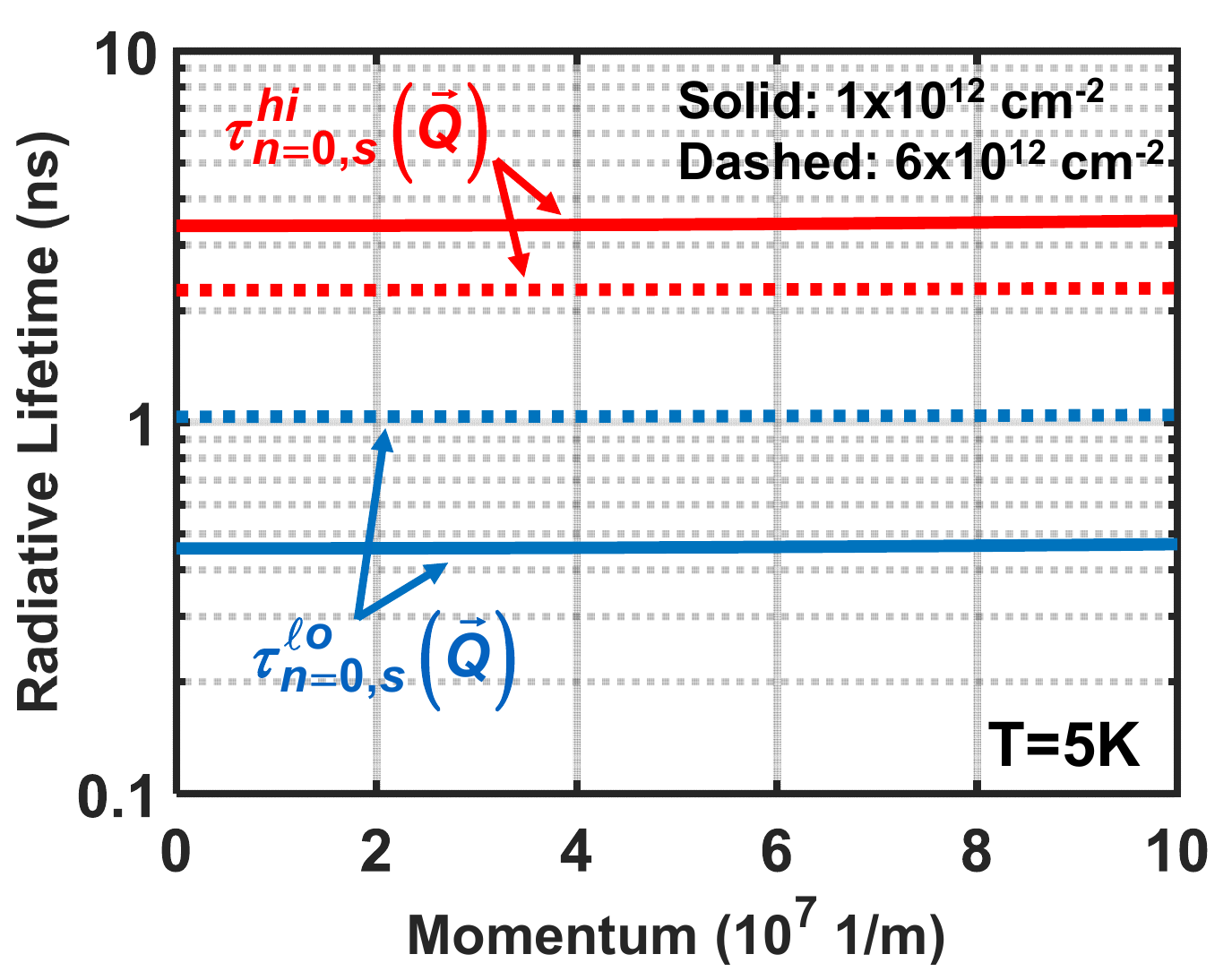}
   \caption{The radiative lifetimes,$\tau^{lo}_{n=0,s}(\vec{Q})$ and $\tau^{hi}_{n-0,s}(\vec{Q})$, of the lower and higher energy eigenstates, respectively, of the coupled exciton-trion system (and corresponding to the lower and higher energy peaks in the optical absorption spectra in Fig.\ref{fig:fig3}) are plotted as a function of the in-plane momentum $Q$ for different electron densities ($10^{12}$ cm$^{-2}$ and $6\times10^{12}$ cm$^{-2}$) for an electron-doped monolayer 2D MoSe$_{2}$ suspended in air. The lifetimes shown correspond to the process depicted in Fig.\ref{fig:fig6}(b) for radiative decay into excited states of the material. T=5 K. The lifetimes shown are three to four orders of magnitude longer than the lifetimes shown earlier in Fig.\ref{fig:fig5} for the process depicted in Fig.\ref{fig:fig6}(a) for radiative decay into the material ground state.}   
    \label{fig:fig7}
  \end{center}
\end{figure}

\begin{figure}[t]
  \begin{center}
   \includegraphics[width=0.9\columnwidth]{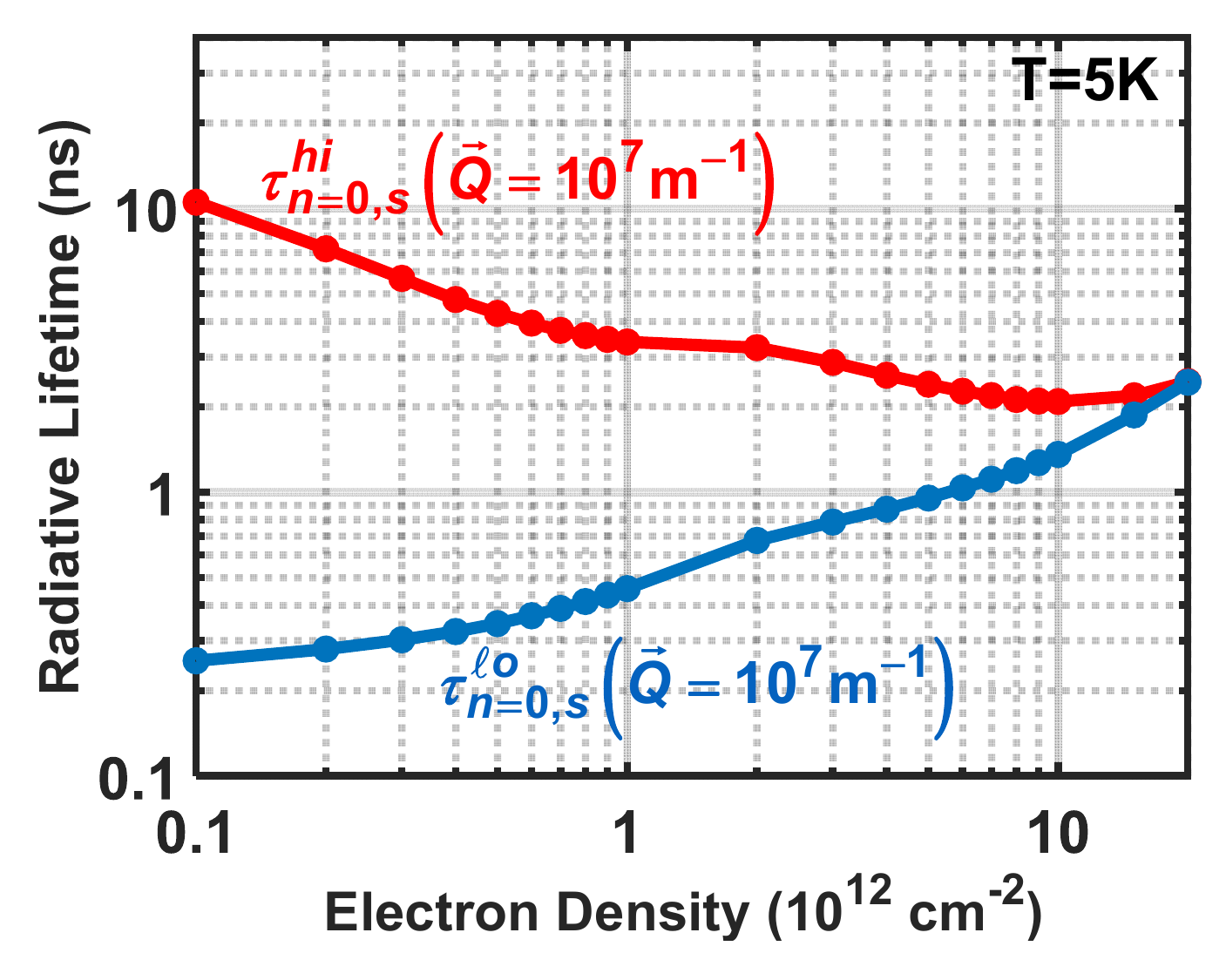}
   \caption{The radiative lifetimes,$\tau^{lo}_{n=0,s}(\vec{Q})$ and $\tau^{hi}_{n-0,s}(\vec{Q})$, of the lower and higher energy eigenstates, respectively, of the coupled exciton-trion system (and corresponding to the lower and higher energy peaks in the optical absorption spectra in Fig.\ref{fig:fig3}) are plotted as a function of the electron densities for an electron-doped monolayer 2D MoSe$_{2}$ suspended in air. T=5 K. The momentum value is chosen to be just outside the light cone $Q\sim 10^{7}$ 1/m. The lifetimes shown correspond to the process depicted in Fig.\ref{fig:fig6}(b) for radiative decay into the excited states of the material.}
    \label{fig:fig8}
  \end{center}
\end{figure}

\subsection{Simulation Results}

Fig.\ref{fig:fig7} shows the radiative lifetimes,$\tau^{lo}_{n=0,s}(\vec{Q})$ and $\tau^{hi}_{n-0,s}(\vec{Q})$, for radiative decay into the excited states of the material, plotted as a function of the in-plane momentum $Q$ of the energy eigenstates for two different electron densities. The radiative lifetimes are finite even outside the light cone and have a weak dependence on the  momentum $Q$. More interestingly, the radiative rates shown in Fig.\ref{fig:fig7} are three to four orders of magnitude smaller compared to the radiative rates for decay into the material ground state shown in Fig.\ref{fig:fig5}. This large difference can be understood as follows. Consider an energy eigenstate of momentum $\vec{Q}$, as given in (\ref{eq:var}), and consider the 4-body bound trion state component of the energy eigenstate (the bound trion state has more weight in the eigenstate than all the unbound trion states). The small radius of the bound trion state ($\sim 1-2$ nm~\cite{Rana20}) means that the phase space occupied by each one of the two CB electrons in the bound trion state is fairly large, and is of the order of $a^{-2}$, where $a$ is the trion radius. When one of the two CB electrons in the bound trion state radiatively recombines with the VB hole, a CB electron and a CB hole are left behind. Suppose the in-plane momentum of the emitted photon is $\vec{Q'}$, the momentum of the CB electron left behind is $\vec{p}+\vec{Q}-\vec{Q'}$, and the momentum of the CB hole is $\vec{p}$. Since $\vec{Q'}$ is restricted to be within the light cone (the phase space area of which is $\sim \omega^{2}/c^{2}$), only a very small portion of the phase space of the CB electron state prior to the photon emission contributes to photon emission. This phase space fraction is of the order of $\omega^{2}a^{2}/c^{2}$, which is between $10^{-3}$ to $10^{-4}$. Note that $\tau^{hi}_{n-0,s}(\vec{Q}) >  \tau^{lo}_{n-0,s}(\vec{Q})$ in Fig.\ref{fig:fig7}, which is the opposite of the case in Fig.\ref{fig:fig5}. This is because the radiative rates in Fig.\ref{fig:fig7} are proportional to $|\beta_{m,s'}|^{2}$ (weight of the trion component in the energy eigenstate), whereas the radiative rates in Fig.\ref{fig:fig5} are proportional to $|\alpha_{n}|^{2}$ (weight of the exciton component in the energy eigenstate). Fig.~\ref{fig:fig8} shows the radiative lifetimes, $\tau^{lo}_{n=0,s}(\vec{Q})$ and $\tau^{hi}_{n-0,s}(\vec{Q})$, for momentum $Q$ value just outside the light cone, plotted for different electron densities. At very small electron densities the radiative lifetime $\tau^{hi}_{n=0,s}(\vec{Q})$ of the higher energy eigenstate is much longer than the lifetime $\tau^{lo}_{n=0,s}(\vec{Q})$ of the lower energy eigenstate, and at very large electron densities these two lifetimes become comparable. The fact that $\tau^{lo}_{n-0,s}(\vec{Q}) << \tau^{hi}_{n-0,s}(\vec{Q})$ at very small electron densities can be understood as follows. At very small electron densities, $|\alpha_{n=0}|^{2} \sim 1$ and $|\beta_{m=0,s'}|^{2} << 1$, and the higher and lower energy eigenstates are thus nearly pure exciton and pure trion states, respectively, and exciton states do not radiatively decay into the excited states of the material.

\begin{figure}[t]
  \begin{center}
   \includegraphics[width=0.9\columnwidth]{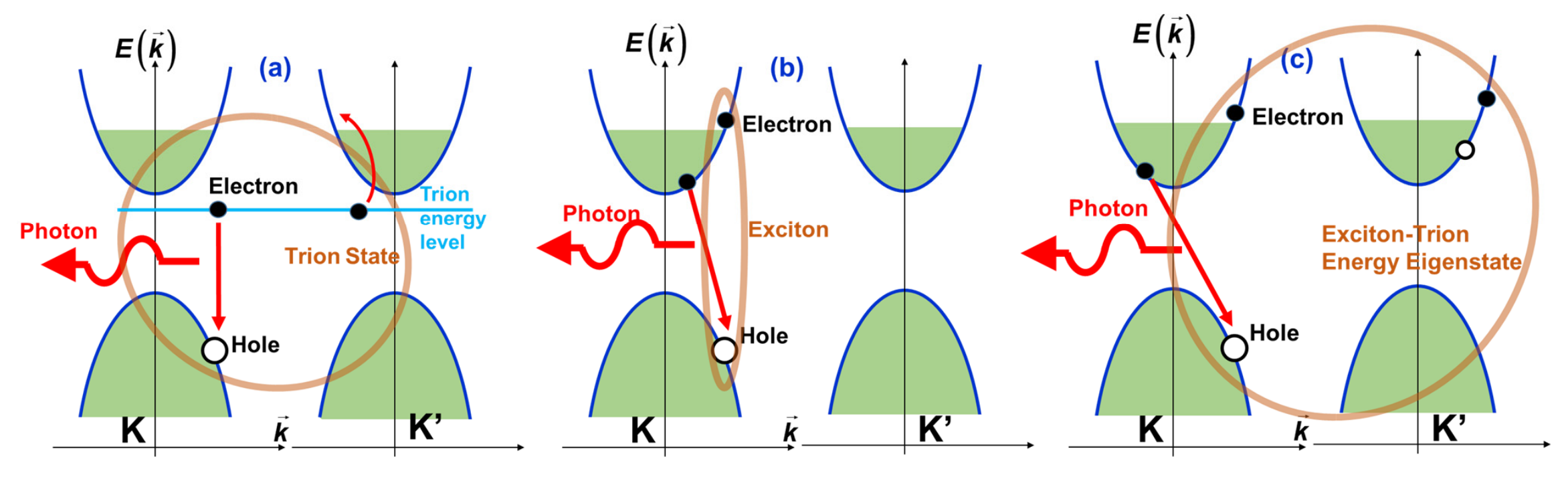}
   \caption{Certain processes that have been proposed in the literature for photon emission involving excitons and trions in electron-doped materials are depicted. (a) Photon emission process involving a 3-body trion state in which the CB electron recombines with the VB hole leaving behind another CB electron which is deposited outside the Fermi sea~\cite{Huard00,Fai13,HWang16}. (b) Photon emission process involving an exciton in which an uncorrelated CB electron from the Fermi sea recombines with the VB hole, leaving behind an electron-hole pair in the CB~\cite{Cotlet20}. (c) Photon emission process involving a trion in which an uncorrelated CB electron from the Fermi sea recombines with the VB hole, leaving behind two electron-hole pairs in the CB~\cite{Cotlet20}.}    
    \label{fig:fig9}
  \end{center}
\end{figure}

\section{Certain Other Misconceptions Regarding Radiative Rates} 
Certain other concepts and processes for radiative transitions have appeared in the literature in the context of excitons and trions in doped 2D materials that are incorrect in the opinion of the authors. We discuss them briefly here. Fig.\ref{fig:fig9}(a) shows a photon emission process involving a 3-body trion state in which the CB electron recombines with the VB hole leaving behind a CB electron which is deposited outside the Fermi sea~\cite{Huard00,Fai13,HWang16}. This model showed that the energy of the photon emitted by a trion state would be red-shifted (with respect to the photon emitted by an exciton in the same material) by roughly the Fermi energy $E_{F}$ (in addition to the trion binding energy) which is consumed in promoting the left-behind CB electron to the unoccupied states above the Fermi level. The red shift of the photon energy with the Fermi energy is in agreement with experiments~\cite{Huard00,Fai13}. However, there are several problems with this photon emission model and with the concept of a 3-body trion state itself~\cite{Rana20}. Recent papers have unambiguously shown that the red-shifting of the lower energy eigenstate, linearly with the Fermi energy, with respect to the higher energy eigenstate is the result of Coulomb interactions~\cite{Rana20,Suris03,Imam16,Macdonald17,Chang19}. Second, this model incorrectly assumes that the electrons forming the trion state are somehow not a part of the CB electronic states (as Fig.\ref{fig:fig9}(a) depicts) and then concludes that the electron left-behind after photon emission needs to be deposited back into the CB with enough energy to avoid Pauli blocking. The closest correct model, depicted in Fig.\ref{fig:fig6}(b), shows that when a 4-body trion state emits a photon, the CB electron and the CB hole left-behind (that were a part of the 4-body trion state) remain in the states they occupied just before the emission of the photon.

Fig.\ref{fig:fig9}(b) shows a photon emission process involving an exciton in which an uncorrelated CB electron from the Fermi sea recombines with the VB hole, leaving behind an electron-hole pair~\cite{Cotlet20}. A simple calculation using an exciton state as the initial state and a final state consisting of a Fermi sea with an electron-hole pair in the CB, and using Fermi's Golden Rule, will show that the rate of this process, although very small, is roughly proportional to the electron density (for small electron densities) which in turn is proportional to the probability of finding an uncorrelated electron near the exciton. The catch here is that the probability of finding an electron of the same spin/valley near the exciton as that of the electron forming the exciton is not proportional to the electron density but is in fact near zero due to Pauli's principle. Each electron in the conduction band, including the one forming an exciton, is surrounded by its exchange hole and the size of this exchange hole is much larger than the size of the exciton in 2D materials for electron densities smaller than $\sim 10^{13}$ cm$^{-3}$. In our model, when we switched from the 4-body operator $T_{\vec{Q}}$ to the connected 4-body operator $T^{c}_{\vec{Q}}$ in (\ref{eq:exciton1}), we removed terms that contributed to the process shown in Fig.\ref{fig:fig9}(b), and one of the difference terms, given in (\ref{eq:conT}), gave the exchange energy contribution, which renormalized the CB energy $E_{c,s}(\vec{k})$ on the LHS in (\ref{eq:exciton1}). The similar process for trions, shown in Fig.\ref{fig:fig9}(c)~\cite{Cotlet20}, would have a negligibly small rate for the same reason.

\section{Discussion and Conclusion}
The results presented in this paper show that photons can be emitted by exciton-trion energy egenstates when their momenta $\vec{Q}$ are inside or outside the light cone. Inside the light cone, radiative rates for transitioning into the material ground state are nearly four orders of magnitude faster than the radiative rates in which the final state is an excited state of the material. Outside the light cone, only radiative decay into an excited state of the material is possible. Our results are expected to clarify many concepts associated with light emission from excitons and trions and their superposition states in doped 2D materials.  

It needs to be mentioned here that the radiative lifetimes measured in experiments depend on the type of measurement performed and therefore some care is needed in comparing experiments with theory. Radiative lifetime measurements are usually performed over exciton/trion ensembles and these ensembles can be prepared in experiments in various ways. Ultrafast resonant optical generation of excitons within the light cone and their subsequent probing via $1s \rightarrow 2s$ excitonic transitions using a mid-IR probe pulse have yielded exciton lifetimes in 2D TMDs that match well with theory~\cite{Huber15}. Time resolved photoluminescence (PL) measurements on the other hand rely on the exciton-trion energy eigenstates to relax down to the light cone before they can recombine radiatively with high efficiency~\cite{Urba16}. This relaxation process is generally bottlenecked by phonon scattering times which are usually much slower (around a few picoseconds) than the radiative lifetimes inside the light cone~\cite{Paras12,Basko16,Ray92,Malic18}. In addition, as discussed in this paper, PL collected from both peaks in the emission/absorption spectra of doped 2D materials are from states that are superpositions of exciton and trion states and contribute to PL from both inside and outside the light cone. Although the radiative rates outside the light cone are much smaller than the rates inside the light cone, the phase space available outside the light cone for hosting a non-equilibrium exciton-trion population is also much larger and a lot more exciton-trions could be present outside the light cone than inside it depending on the nature and details of the experiment. An accurate modeling of radiative emission from non-equilibrium ensembles requires computational approaches well beyond the scope of this work~\cite{Malic18}.

\section{Acknowledgments}
The authors would like to acknowledge helpful discussions with Nick Vamivakas, Francesco Monticone, and Jacob Khurgin, and support from CCMR under NSF-MRSEC grant number DMR-1719875 and NSF EFRI-NewLaw under grant number 1741694.


\begin{thebibliography}{99}
\bibitem{Rana20b} In principle, 4-body trion states can have a non-zero optical matrix element with the ground state if one considers, for example, a statistical mixture ground state at a non-zero temperature with thermally excited electron-hole excitations across the Fermi energy. But even then, the direct contribution of 4-body trion states to the material optical conductivity is found to be very small.
\end{thebibliography}
\end{document}